\documentclass[runningheads]{llncs}

\usepackage[T1]{fontenc}
\usepackage{graphicx}
\usepackage{float}
\usepackage{xcolor}
\usepackage{url}
\usepackage{tabularx}
\usepackage{makecell}
\usepackage{hyperref}

\begin{document}

\title{Can GPT-4o Evaluate Usability Like Human Experts? A Comparative Study on Issue Identification in Heuristic Evaluation}

\titlerunning{Can GPT-4o Evaluate Usability Like Human Experts?}

\author{Guilherme Guerino\inst{1} \and Luiz Rodrigues\inst{2} \and Bruna Capeleti\inst{3} \and Rafael Ferreira Mello\inst{4} \and André Freire\inst{3} \and Luciana Zaina\inst{5}}

\authorrunning{Guerino et al.}

\institute{State University of Paraná, Apucarana, Brazil\\\email{guilherme.guerino@ies.unespar.edu.br} 
\and Federal Technological University of Paraná, Apucarana, Brazil \\\email{luizrodrigues@utfpr.edu.br}
\and Federal University of Lavras, Lavras, Brazil\\\email{\{brunacapeleti,apfreire\}@gmail.com}
\and Federal Rural University of Pernambuco, Recife, Brazil\\\email{rafael.mello@ufrpe.br}
\and Federal University of São Carlos, Sorocaba, Brazil\\\email{lzaina@ufscar.br}}

\maketitle             

\vspace{-1cm}
\begin{abstract}
Heuristic evaluation is a widely used method in Human-Computer Interaction (HCI) to inspect interfaces and identify issues based on heuristics. Recently, Large Language Models (LLMs), such as GPT-4o, have been applied in HCI to assist in persona creation, the ideation process, and the analysis of semi-structured interviews. However, considering the need to understand heuristics and the high degree of abstraction required to evaluate them, LLMs may have difficulty conducting heuristic evaluation. However, prior research has not investigated GPT-4o's performance in heuristic evaluation compared to HCI experts in web-based systems. In this context, this study aims to compare the results of a heuristic evaluation performed by GPT-4o and human experts. To this end, we selected a set of screenshots from a web system and asked GPT-4o to perform a heuristic evaluation based on Nielsen's Heuristics from a literature-grounded prompt. Our results indicate that only 21.2\% of the issues identified by human experts were also identified by GPT-4o, despite it found 27 new issues. We also found that GPT-4o performed better for heuristics related to aesthetic and minimalist design and match between system and real world, whereas it has difficulty identifying issues in heuristics related to flexibility, control, and user efficiency. Additionally, we noticed that GPT-4o generated several false positives due to hallucinations and attempts to predict issues. Finally, we highlight five takeaways for the conscious use of GPT-4o in heuristic evaluations.

\end{abstract}

\section{Introduction}

Assessment of User Interfaces (UI) is crucial in developing software. Usually, this assessment focuses on evaluating UI and user interaction using various criteria, such as usability \cite{tullis2013}, which is defined as \textit{"the degree to which a system, product, or service allows specified users to achieve specific goals with efficiency, effectiveness, and satisfaction in a specified context of use}" \cite{ISO9241-11:2018}. Usability plays an important role in the acceptance of software by users, and therefore it becomes essential to carry out usability evaluations \cite{tullis2013,moran2019}. Although usability testing is a valuable method for collecting feedback from real users, it can require more time and costs to perform \cite{tullis2013,stephanidis2024}. In this sense, usability inspections can be employed as an alternative. They involve HCI experts reviewing the UI to identify issues in the interface and its available interaction with the software \cite{stephanidis2024}.

Usability inspections can be performed at different stages of software development or even as a predictive evaluation before user testing. Heuristic evaluation is a well-known method commonly adopted for usability evaluation \cite{stephanidis2024,nielsen1994heuristic}. To conduct a heuristic evaluation, Human-Computer Interaction (HCI) experts are provided with a list of heuristics (i.e., a set of principles) and are tasked to review the interface design and note down any issues violating the provided heuristics \cite{tullis2013,nielsen1994heuristic}. Many heuristics have been created with different purposes and applications, e.g. accessible mobile \cite{mi2014}, touchscreen and multimodal \cite{gorlewicz2020,Konopatzki2024}, and social network \cite{saavedra2019}. However, Nielsen's 10 heuristics are seen as the most consolidated and recognized for usability inspection \cite{nielsen1994heuristic}. Typically, heuristic evaluation is carried out by multiple experts since it involves human judgment \cite{stephanidis2024}. The multiple viewpoints allow a comparison of findings, thereby reducing the potential bias inherent in relying on a single evaluator.

Recently, Large Language Models (LLMs), such as GPT, Claude, Llama, among others, have gained popularity for a range of human activities (e.g., software engineering, documents elaboration, and wireframe elaboration) \cite{wang2024history}. These models are based on artificial intelligence techniques to enable users to interact with them using natural language \cite{ziyu-etal-2023-lens}. The literature has investigated the role of LLMs in supporting HCI activities, including UI sketching and UI testing and inspection \cite{takaffoli2024}. In the scope of heuristic evaluation, studies leverage LLMs' capability to emulate human cognitive processes, potentially offering feedback on how each UI element follows or violates a set of heuristics \cite{duan2024}. However, LLMs still face challenges and restrictions with regard to their output for activities that involve a high level of abstraction \cite{duan2024}. In heuristic evaluation, LLMs might struggle due to the nuanced interpretation required for understanding heuristics, which they have yet to fully achieve \cite{duan2024,stephanidis2024}. This raises questions about the connection between knowledge and content produced by these models \cite{wang2024history}. 

Motivated by these challenges and limitations, this paper presents an investigation comparing the results of a heuristic inspection carried out by human experts and GPT-4o, the main LLM able to interpret images at this research's time \cite{openai2023gpt4}. To carry out the study, a set of user interfaces was selected from a web-based system that integrates, systematizes and disseminates information based on maps and graphs about mineral research and production, which was inspected using Nielsen's heuristics. The results showed that only 21.2\% of the issues identified by the experts were also identified by GPT-4o. Furthermore, although there was no significant difference between the severities of the issues identified by both groups, GPT-4o presented better results for heuristics directly linked to interface design, while experts excelled in heuristics related to interaction. Besides, we highlight that 24.3\% of the issues identified by GPT-4o were classified as false positives, indicating that the tool can hallucinate or try to predict issues that do not exist wrongly. 

As contributions from this study, we can highlight the description of a complete process of classification of issues identified by GPT-4o, including classification of false positives; a set of five takeaways to assist the conscious use of GPT-4o in heuristic evaluations; and the creation, experimentation, and provision of a prompt that can be used and refined by researchers and HCI professionals in heuristic evaluations with LLMs.

\vspace{-0.2cm}
\section{Theoretical Background}

This section presents fundamentals on heuristic evaluation and LLMs, and discusses the related work and their differences and similarities with this study.

\vspace{-0.3cm}
\subsection{Heuristic Evaluation}

Heuristic evaluation consists of an usability inspection method that aims to identify usability issues based on established heuristics or principles. Originally introduced by Nielsen and Molich \cite{molich1990improving}, this method involves a systematic assessment by usability experts who analyze a product to detect potential usability issues. In the traditional form, experts conduct and document the identified issues alone, then the evaluators consolidate their findings into a unified list. Lastly, the issues are classified according to their frequency and severity \cite{nielsen1994heuristic}. There is another approach, focused to reduce the time and avoid long discussions, where the experts detect the issues together and then assess the severity levels independently \cite{petrie2010collaborative}. The severity levels that can be related to a mapped issue are: 0 for not a issue, 1 for a cosmetic issue, 2 for minor issue, 3 for major issue and 4 for catastrophic issue. The higher the severity level, the faster should be the process to fix the issue.

Nielsen \cite{nielsen1994heuristic} proposed a set of 10 heuristics that are used primarily as a guide in usability evaluations: 1) visibility of system status, 2) match between system and the real world, 3) user control and freedom, 4) consistency and standards, 5) error prevention, 6) recognition rather than recall, 7) flexibility and efficiency of use, 8) aesthetic and minimalist design, 9) help users recognize, diagnose, and recover from error, and 10) help and documentation. In our work, we used these heuristics for evaluation.

\vspace{-0.3cm}
\subsection{Large Language Models (LLM)}

LLMs are trained using large datasets of text from diverse sources, such as books, websites, and articles, allowing them to learn patterns, structure, and semantics of human language. The training involves processing this data through a neural network, typically based on the transformer architecture, which excels at handling sequential data \cite{Maldonado13370}. During training, the model learns to predict the next word in a sentence given the previous context, building an understanding of language through exposure to many examples. This process requires significant computational resources and often spans weeks or months, continually refining the model's ability to generate coherent and contextually appropriate text \cite{wang2024history}. The outcome is a model capable of performing a wide range of language tasks, from simple sentence completion to complex content synthesis \cite{chang2024survey}.

LLMs were firstly designed to make sense of human language and generate text that sounds natural and relevant. These models help with tasks like automating customer chats, creating written content, answering educational questions, and analyzing data \cite{ziyu-etal-2023-lens,rodrigues2024assessing}. Furthermore, LLMs like GPT-4o not only handle text input, but they also interpret images \cite{openai2023gpt4}. For instance, one might prompt it with an image along with a textual command to receive a textual answer such as transcription of the image \cite{pereira2024vlm}. Despite these advancements, generating the expected content is not trivial, where one must effectively prompt the model to increase its chances of creating relevant content \cite{karmaker-santu-feng-2023-teler}. 

In that context, \textit{prompt engineering} is essential for achieving the best results from LLMs. By carefully crafting the questions or commands given to these models, users can guide them to produce the more useful responses \cite{wei2022chain}. However, creating effective prompts can be challenging, requiring a solid understanding of the writing aspects that are more likely to lead the model towards the expected output and, hence, achieve accurate and unbiased outcomes \cite{karmaker-santu-feng-2023-teler}. 

Although challenging, using LLMs continue to be explored in academic and industrial domains \cite{chang2024survey}. Particularly, LLMs in HCI can improve the design, development, and evaluation of user interfaces and experiences. One application is the extraction of user needs from interviews. For example, \cite{freitas2024sensation} conducted a usability study followed by interviews, then used GPT-4o to perform a thematic analysis on the interviews's transcripts. Another application is image generation. LLMs like GPT-4o are able to generate images from textual prompts, which might be used to support interface design by providing the model with specifications and asking it to generate interface suggestions \cite{brade2023promptify}. Accordingly, LLMs are valuable tools that can help HCI research and development from various perspectives.

\vspace{-0.3cm}
\subsection{Related Work}

The use of LLMs, such as GPT-4o, has been explored in HCI research topics. Several studies have used LLMs to generate synthetic data \cite{de2023can,tavast2022language} or participants \cite{hamalainen2023evaluating}, explore and interpret user personas \cite{liu2024personaflow,barambones2024chatgpt,panda2024llms}, and assist in the brainstorming and ideation process \cite{girotra2023ideas,suh2024luminate}. Other studies have used LLMs to assist in creating interfaces through prompts \cite{lu2022bridging,petridis2023promptinfuser} and make the prototyping process more efficient in development teams \cite{leiker2023prototyping}.

Some studies have also explored the use of LLMs in heuristic evaluations. Zhong et al. \cite{zhongcan} explored the reliability of LLMs in performing heuristic evaluations on two mobile applications, comparing their results with those of human evaluators (research assistants). The results indicate that LLMs can generate and describe usability issues in the same way as human evaluators. The authors emphasize that more studies are needed to produce reliable results \cite{zhongcan}.

Duan et al. \cite{duan2024} developed a plugin for Figma that allows designers to evaluate mobile interfaces and receive text feedback automatically. The authors conducted tests with expert designers and found that the plugin based on GPT-4o performs well for poor UIs but that its usefulness decreases as the UI improves. Furthermore, the authors stated that, despite some limitations, GPT-4o can evaluate some heuristics, while others require further technical advances  \cite{duan2024}.

Moreover, Meinecke et al. \cite{meinecke2024comparative} compared the results of a heuristic evaluation performed by GPT-4o with the results of seven human inspectors without experience in heuristic evaluation. The human evaluators had access to a demo version of the system to be evaluated, while GPT-4o evaluated the HTML code of the same application. As a result, the authors highlight that GPT-4o could not identify usability issues better than human evaluators, stating that GPT-4o can act as an assistant to the evaluators. Finally, the authors highlight that more research is needed to verify how GPT-4o can assist human evaluators \cite{meinecke2024comparative}.

Considering the works described above, they investigated mainly the UI of mobile applications or compared the results of GPT-4o with those of novices and research assistants. Our study, unlike others, uses GPT-4o to evaluate a web-based system to further contribute to research in this area. It compares the results with three human experts who performed the evaluation using the same heuristics at a previous time. Furthermore, unlike other studies, we analyzed the issues in terms of agreement in the description, heuristics, severity, issues identified only by GPT-4o, and false positives to propose takeaways that can help researchers and professionals who want to perform a heuristic evaluation with GPT-4o or another LLM. Finally, we developed, tested, and presented a literature-grounded prompt with prompt engineering elements such as instruction, role, context, think, step by step, rubric, justification, and output.

\section{Method}

This study compared the outcomes of a heuristic evaluation performed by human experts with those produced by GPT-4o on the same system. In this context, we aim to address the following research questions (RQ): (RQ1) \textit{What is the agreement between heuristic evaluation results by experts and those by GPT-4o, in terms of issue compliance, violated heuristics, and severity?}; (RQ2)\textit{ What are the characteristics of the issues found exclusively by GPT-4o?}; (RQ3) \textit{What are the characteristics of issues considered false positives identified by GPT-4o?}

We used the guidelines provided by Lazar et al. \cite{lazar2017research} to define the steps of the methodology in planning, execution, and analysis of our comparative study. Besides, we used a mixed analysis method to verify two output variables: (i) agreement in identifying issues and (ii) the characteristics of the issues and false positives identified by GPT-4o.

\subsection{Planning}

Our study compared the results of a previously published heuristic evaluation performed by human experts \cite{capeleti2023} with the results collected from GPT-4o. In the previous study \cite{capeleti2023}, three experts (two with PhDs in the field of HCI and one master's student who was also a specialist in HCI) conducted a collaborative heuristic evaluation, and a discussion was performed to agree on the severity level. We emphasize that, for the comparison, we used the results achieved by the authors of this previously published study \cite{capeleti2023}, which highlight the entire review and validation process of the defects found by these human experts. The system evaluated was the P3M\footnote{https://p3m.sgb.gov.br}, a public platform to subsidize studies and research focused on mineral production. P3M's data and information collection includes geoscientific, technical-economic, regulatory, socioeconomic, and socio-environmental aspects related to the mineral industry's chain of activities, including mineral research, mining development, mining, and mineral transformation. The human experts identified 66 issues in that system, which served as the basis for our comparison (i.e., human experts data-related).

To conduct the heuristic evaluation from the LLM perspective, we chose the Open AI gpt-4o-2024-08-06 model, which, at the time of this study (November 2024), represents one of the main LLMs for answering textual questions based on a given image and was accessed through the Open AI playground interface \footnote{https://platform.openai.com/docs/overview}. The prompt engineering technique \cite{chen2023unleashing,sahoo2024systematic} supported the elaboration and structure of the prompt applied in our study. Before performing the GPT-4o evaluation, we carried out a pilot test with two screenshots. In this context, we tested the screenshots as the prompt's ability to perform the evaluation and identify issues. In the pilot test, 11 issues were identified in the two screenshots, each defining its usability issue, the violated heuristic, the severity, the explanation, and the position on the screen. All researchers involved in the study considered that the pilot study's prompt was adequate for the goal. Table \ref{tab:prompt}, for reasons of space, shows the elements used and their definition (based on works about prompt engineering \cite{mello2024prompt,freitas2024sensation,rodrigues2024assessing,white2023prompt,giray2023prompt}, and a summary of the text used in the prompt for each element. The complete prompt can be accessed at this link\footnote{http://doi.org/10.6084/m9.figshare.28400078}
.

\vspace{-0.3cm}
\subsection{Execution}

We selected 20 system screenshots for heuristic evaluation with GPT-4o (see the link\footnote{https://doi.org/10.6084/m9.figshare.28373342})
. These 20 screenshots were chosen by mutual agreement between the paper's authors, considering they covered most system functionalities. Thus, the prompt shown in Table \ref{tab:prompt} was executed once (one-shot) for each screenshot. We highlight that the screenshots used in the pilot test were not included in the final study. As the "Output" element of the prompt shows, each prompt running generated a CSV spreadsheet with the issues identified in that screenshot. In the end, a single spreadsheet was created to facilitate the analysis by compiling all the partial spreadsheets.

\vspace{-0.4cm}
\begin{table}[ht]
    \centering
    \caption{Summary of the prompt used in GPT-4o to perform heuristic evaluation.}
    \scriptsize
    \begin{tabular}{lp{4cm}p{6.2cm}}\hline
     \textbf{Element} & \textbf{Description} &\textbf{Prompt}\\\hline
     Instruction & The task the LLM is expected to do & Perform a heuristic evaluation of the attached interface based on Jakob Nielsen's 10 general principles for interaction design: 1) Visibility of System Status (The design should always keep users informed about what is going on, through [...]\\
     Role & A role model the LLM is expected to assume & You are an industry designer with experience in User Experience and User Interface.\\
     Context & The setting to consider in answering the prompt & You are inspecting the interface of an interactive website.\\
     Think & A guideline for the model to \textit{think} before answering the prompt & Think step by step\\
     Step by step & Detailed guidelines on how the LLM should \textit{think} & 1. Inspect the interface to get a feel for the flow of the interaction and the general scope of the system. 2. To focus on specific interface elements while knowing how they fit into the larger whole. 3. Elaborate the rationale to justify [...]\\
     Rubric & Criteria for the LLM to consider in generating its output & It is not sufficient for evaluators to simply say that they do not like something; they should explain why they do not like it with reference to the heuristics or to other usability results. The evaluators should try to be as specific as possible and [...]\\
     Justification & A guideline for the LLM to reason about the output it is generating before outputing it & Reasoning about the justification for your response by explaining why you made the choices you actually made.\\
     Output & Guidelines on how the output should look like & Your answer must be formatted as a CSV where there is a row for each usability problem identified and nothing else. Each row must include the usability problem, the usability heuristics it violates, the problem severity (1 to 4), a detailed [...]\\\hline
     \multicolumn{3}{l}{\textbf{See the complete prompt in \url{https://doi.org/10.6084/m9.figshare.28400078}
     }}\\
    \end{tabular}
    \label{tab:prompt}
\end{table}
\vspace{-0.6cm}

The classification of duplicates, false positives, or consolidated issues of all issues identified by GPT-4o was based on the interpretation and evaluation of four researchers (R1, R2, R3, and R4) involved in this paper. All researchers have extensive experience in HCI research, one with a master's degree (R3) and three with a PhD in the area (R1, R2, and R4). In the first moment, R1 analyzed all issues identified by GPT-4o, grouping issues in duplication (i.e., issues identified two or more times in different screenshots) and then summarizing them in only one issue (i.e., duplicated issues were considered only once in final sample). R2 reviewed the grouping made by R1 and suggested new groupings. R1 and R2 then resolved the disagreements.

To proceed, R1 analyzed each remaining issue to verify whether it was a issue or a false positive. The following were considered false positives: i) problem disagreements, when the researcher disagrees with GPT-4o that the problem pointed out is a issue; ii) problem assumptions, when a issue pointed out by GPT-4o is impossible to detect in screenshots, for example, interaction issues on static screens; and iii) generalized problems, when the issue is very generic and does not explicitly point out where the problem is. After R1's analysis, R2 reviewed the consolidation and noted comments, and at the end R1 and R2 solved the disagreements. It is worth noting that the researchers did not consider the severities information in the analysis of the issues. Next, R1 analyzed each remaining issues identified by GPT-4o and compared it with the issues identified by the human experts. R3 then reviewed the results obtained by R1, pointing out its disagreements. Finally, R2 and R4 consolidated the results, resolving the disagreements pointed out by R2.

\vspace{-0.3cm}
\subsection{Data Analysis}

With the classification consolidated by the four researchers, we conducted a mixed-method analysis to address our research questions. For RQ1, we employed descriptive analysis as well as Pearson's chi-square test \cite{pearson1900x} to assess the agreement between the issues identified by experts and those found by GPT-4o. This test is widely used to assess associations between categorical variables, making it suitable for analyzing whether the distribution of issues identified by GPT-4o and humans experts occurs independently or follows a specific pattern. Since the issues were classified into distinct categories without a continuous numerical scale, the Pearson's chi-square test \cite{pearson1900x} allows us to determine whether there is a statistically significant dependence between the groups, i.e., whether GPT-4o and human experts tend to identify the same types of issues or if their analyses follow different patterns. 

Additionally, we used the Mann-Whitney test \cite{mcknight2010mann} to analyze the agreement concerning the severity of the issues identified by the groups. This test is appropriate for comparing the distributions of two independent samples when the data are ordinal and do not follow a normal distribution. Our study classified issue severity into levels, e.g., minor or major issues. The Mann-Whitney test \cite{mcknight2010mann} helps determine whether there are significant differences between the severity ratings assigned by human experts and GPT-4o, identifying whether one group tends to classify issues as more or less severe compared to the other.

Finally, for RQ2 and RQ3, we carried out a thematic analysis \cite{braun2024thematic} to identify the consolidated issues and false positives characteristics identified by GPT-4o.

\vspace{-0.2cm}
\section{Results}
In this section, we first present an overview of the quantitative results obtained in the consolidation process of issues identified by GPT-4o. Second, we present the agreement results of issues identified by GPT-4o and human experts (RQ1). Then, we show the characteristics of consolidated issues identified only by GPT-4o (RQ2) and, finally, the characteristics of false positives (RQ3).

\vspace{-0.3cm}
\subsection{Overview}

Using the 20 screenshots and the prompt shown in Table \ref{tab:prompt}, GPT-4o identified a total of N = 111 issues (mean=5.55, SD=1.14). Of these issues, 43 (38.7\%) were considered duplicates. Of the remaining 68 issues, 27 (24.3\%) were considered false positives. The sample of consolidated issues was N = 41 (36.9\%), showed in Table \ref{tab:issues}. We make available all issues by screenshot at the link\footnote{https://doi.org/10.6084/m9.figshare.28373354}
, and the classification of duplicates, consolidated issues and false positives at the link\footnote{https://doi.org/10.6084/m9.figshare.28373363}
.

\vspace{-0.5cm}
\begin{table}[ht]
    \centering
    \caption{Consolidated issues identified by GPT-4o in the heuristic evaluation.}
    \scriptsize
    \begin{tabular}{clcc}\hline
     \textbf{ID} & \textbf{Description} & \textbf{Heuristic} & \textbf{Severity} \\\hline
     D1 & Information overload in the side panel & H8 & 2\\
     D2 & Lack of clarity in the function of zoom icons & H2 & 3\\
     D3 & Uninformative "Home" page title & H6 & 2\\
     D4 & Lack of clear indication of active selection & H3 & 3\\
     D5 & Problem in accessing language settings intuitively & H2 & 2\\
     D6 & Difficulty identifying icons in the side menu & H2 & 3\\
     D7 & Difficulty accessing support or help directly & H10 & 2\\
     D8 & Map without clear legend & H5, H2 & 3 \\
     D9 & Lack of option to undo actions & H3 & 3\\
     D10 & Inadequate visibility of system status & H1 & 3\\
     D11 & Interface elements combined in a confusing way & H2 & 2\\
     D12 & Information overlapping on the map & H8 & 2\\
     D13 & Lack of contrast between important elements & H8 & 3\\
     D14 & Use of technical terms without explanation & H2 & 2\\
     D15 & Dense side menu without clear organization & H4 & 2\\
     D16 & Small size of zoom buttons & H7 & 2\\
     D17 & Choice of low-contrast colors for menu icons & H8 & 2\\
     D18 & Lack of clarity in navigation between categories & H2 & 3\\
     D19 & Button names are not self-explanatory & H6 & 2\\
     D20 & Status visibility problem & H1 & 2\\
     D21 & Aesthetic and layout problem & H8 & 1\\
     D22 & Potentially irrelevant information & H8 & 2\\
     D23 & System status visibility & H1 & 3\\
     D24 & Insufficient navigation text & H6 & 3 \\
     D25 & Lack of action confirmation & H5 & 3\\
     D26 & Search fields no consistent & H4 & 3\\
     D27 & Ambiguous "No active layers!" message & H5 & 2\\
     D28 & Navigation between features is not clear & H3 & 3\\
     D29 & Inadequate visibility of filter status & H1 & 3\\
     D30 & Slow loading of maps & H1 & 4\\
     D31 & Overlapping of features & H8 & 2\\
     D32 & Lack of coordinate indication & H6 & 2\\
     D33 & "Road" label in English & H2 & 2\\
     D34 & Close button design & H1 & 3\\
     D35 & Visual index of maps & H6 & 2\\
     D36 & Unclear "Clear all" button & H5 & 3\\
     D37 & Unclear unread badges & H1 & 2\\
     D38 & Lack of icon customization & H7 & 2\\
     D39 & Lack of system status indication when drawing & H1 & 3\\
     D40 & Basemaps with no context description & H2 & 2\\
     D41 & Insufficiently highlighted coordinate information & H1 & 1\\\hline
    \end{tabular}
    \label{tab:issues}
\end{table}

\vspace{-0.5cm}
Looking at Table \ref{tab:issues}, we verified that GPT-4o identified issues related to navigation between functionalities (e.g., D3 and D27), visibility and feedback (e.g., D28 and D39), interface design (e.g., D1, D15, and D17), lack of clarity of contextual information (e.g., D26, D14, and D35), lack of customization (e.g., D9 and D25), and support difficulties (e.g., D5 and D14). 

\subsection{Agreement of results (RQ1)}

In the previous work \cite{capeleti2023}, the human experts found 66 defects in the system. Of these issues, N = 14 (21.2\%) were also found by GPT-4o. Table \ref{tab:both} shows the issues that were found by both groups and their description, heuristic violated for that issue and its severity. 

\vspace{-0.5cm}
\begin{table}[ht]
    \centering
    \caption{Issues identified by both GTP-4 and experts.}
    \scriptsize
    \begin{tabular}{cp{4cm}ccp{0.2cm}p{4cm}cc}\hline
  &  \textbf{GPT-4o} & && & \textbf{EXPERTS} & &\\\hline
     \textbf{ID} & \textbf{Description} & \textbf{H*} & \textbf{S*} && \textbf{Description} & \textbf{H*} & \textbf{S*}\\\hline
     D3 & Uninformative "Home" page title & H6 & 2 && "Home" link - what does it mean? & H6 & 3\\
     D4 & Lack of clear indication of active selection & H3 & 3 && There is no clear correspondence between colors and indicator types when there are many being selected & H1 & 3\\
     D6 & Difficulty identifying icons in the side menu & H2 & 3 && Icon at the top left is not actually a menu & H4, H7 & 3\\
     D7 & Difficulty accessing support or help directly & H10 & 2 && There is no resource to explain what the percentage is & H6 & 1\\
     D8 & Map without clear caption & H5, H2 & 3 && Captions - description is not formatted according to standard & H6 & 1\\
     D12 & Information overlapping on the map & H8 & 2& & Filter modal that is placed on top of a data column, hindering the visualization & H7 & 3\\
     D13 & Lack of contrast between important elements & H8 & 3 && Colors of charts to relate to indicators are difficult to identify (thin bar) & H3, H7 & 2\\
     D14 & Use of technical terms without explanation & H2 & 2 && There is no explanation for specific technical terms for lay audiences & H10 & 3\\
     D15 & Dense side menu without clear organization & H4 & 2 && Filters by region/state do not make sense & H2 & 4\\
     D17 & Choice of low-contrast colors for menu icons & H8 & 2 && Information is identified only by color - this may make it difficult for other users to understand & H2, H4, H6 & 2\\
     D20 & Status visibility problem & H1 & 2 && Layer list - it was not clear what are active layers and the layers per theme that can be activated & H1, H6 & 3\\
     D25 & Lack of action confirmation & H5 & 3 && Favorite option - only favorite main layer? What can be favorited? & H3, H6, H7 & 3\\
     D26 & Search fields no consistent & H4 & 3 && Why is there a country filter if there is only Brazil & H5, H6 & 3\\
     D37 & Unclear unread badges & H1 & 2 && Number along with layers has mobile notification affordance - it gives the impression that when you click the number it will disappear & H4, H6 & 2\\\hline
     &\textbf{H* - Heuristic; S* - Severity.}&&&&&\\
    \end{tabular}
    \label{tab:both}
\end{table}
\vspace{-0.5cm}

Although the 14 issues were identified by both groups, by analyzing Table \ref{tab:both}, we verified that the writing of the description differs, with the experts being more descriptive and provocative, sometimes even using questions to describe the issue (e.g., D25), while GPT-4o is more direct and succinct (e.g., D26). By applying Pearson's chi-square test \cite{pearson1900x} to the sample of all issues and the identification or not by the groups, we yielded significant results ({${\chi}^2$(1)= 34.1, p-value < .001}), indicating that there is a statistically significant difference between the identification of issues by experts and by GPT-4o. Table \ref{tab:chi} shows the contingency table obtained with the test, where "yes" means the group identified a problem, and "no" means the group did not identify a problem.

Table \ref{tab:heuristics} shows the percentage of issues per heuristic for each group (i.e., GPT-4o and human experts). When the number of issues identified by each heuristic of the expert group is added together, the number is much higher than the total number of issues identified by the group (N = 66). This result occurs because 37.8\% of the issues (N = 25) violated two or more heuristics in the expert group. 

\begin{table}[htb]
    \begin{minipage}{.48\linewidth}
      \caption{Pearson's chi-square test \\ contingency table.}
        \begin{tabular}{ccccc}\hline
         &&\multicolumn{2}{c}{\textbf{GPT-4o}}& \\\hline
        \textbf{Experts}&&\textbf{Yes}&\textbf{No}&\textbf{Total}\\\hline
         \textbf{Yes} & Observed & 28 & 52 & 80\\
             & Expected & 41.1 & 38.9 & 80.0\\  
         \textbf{No} & Observed & 27 & 0 & 27\\
             & Expected & 13.9 & 13.1 & 27.0\\
        \textbf{Total} & Observed & 55 & 52 & 27\\
             & Expected & 55.0 & 52.0 & 27.0\\\hline
            &\textbf{Value}&\textbf{gl}&\textbf{p}&\\\hline
         \textbf{$\chi^2$} & 34.1 & 1 & < .001 & \\
         \textbf{N} & 107 & & &\\\hline
    \end{tabular}
    \label{tab:chi}
    \end{minipage}
    \begin{minipage}{.5\linewidth}
        \caption{Number of issues that violated each heuristic.}
        \begin{tabular}{ccccc}\hline
         & \textbf{GTP-4} & \% & \textbf{Experts} & \%\\\hline
         \textbf{Heuristics} & \textbf{41} & \textbf{100\%} & \textbf{66} & 100\%\\\hline
         H1 & 9 & 21.9\% & 15 & 22.7\%\\
         H2 & 9 & 21.9\% & 6 & 9.0\%\\
         H3 & 3 & 7.3\% & 13 & 19.7\%\\
         H4 & 2 & 4.8\% & 5 & 7.5\%\\
         H5 & 4 & 9.7\% & 8 & 12.1\%\\
         H6 & 5 & 12.2\% & 31 & 46.9\%\\
         H7 & 2 & 4.8\% & 14 & 21.2\%\\
         H8 & 7 & 17.0\% & 5 & 7.5\%\\
         H9 & 0 & 0.0\% & 2 & 3.0\%\\
         H10 & 1 & 2.4\% & 1 & 1.5\%\\\hline
        \end{tabular}
         \label{tab:heuristics}
    \end{minipage} 
\end{table}
\vspace{-0.5cm}

When observing the results of GTP-4, we noticed that only one issue (D8, see Table \ref{tab:issues}) violated two heuristics. Then, when analyzing and comparing the percentages of the total issues in each group, we noticed that GPT-4o stood out more significantly among the experts in H2 (compatibility between the system and the real world) and H7 (aesthetics and minimalist design). On the other hand, the experts were able to identify more issues than GPT-4o for H3 (control and freedom for the user), H6 (recognition instead of memorization), and H7 (efficiency and flexibility of use). Interestingly, we noticed that H9 (help users recognize, diagnose, and recover from errors) and H10 (help and documentation) allowed the identification of a few issues for both groups. We highlight that H9 was the only heuristic not violated by any issue identified by GPT-4o.

\begin{figure}[ht]
    \centering
    \fbox{\includegraphics[width=.98\linewidth]{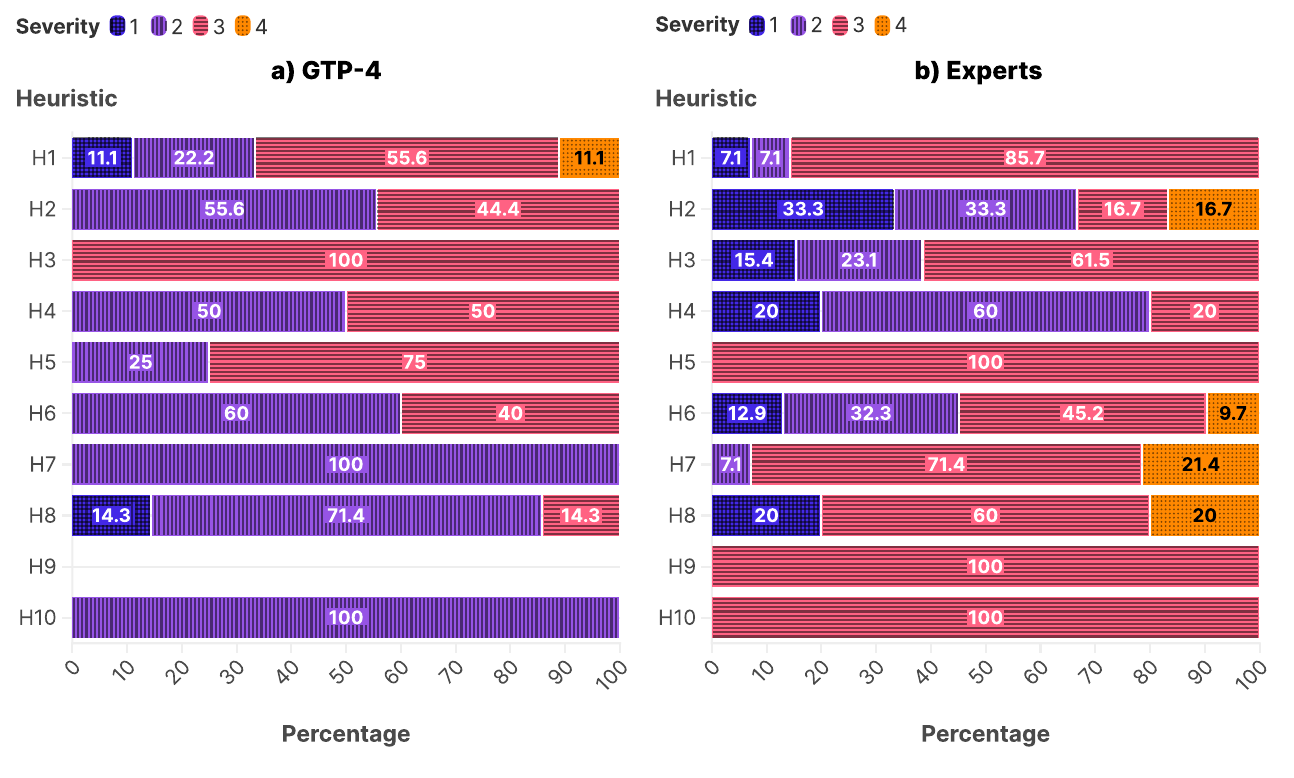}}
    \caption{Severity of issues identified by heuristics.}
    \label{fig:heuxsev}
\end{figure}

Considering the severity of the defects (see Figure \ref{fig:heuxsev}), we observed that both groups were likelier to find issues of severity 2 (minor usability issue) or 3 (major usability issue). The only issue of severity 4 (catastrophic) identified by GPT-4o was D30 (see Table \ref{tab:issues}), which concerns the slow loading of the system maps. When applying the Mann-Whitney U test \cite{mcknight2010mann} to compare the differences in issue severities identified by each group, we found that U($N_{Experts}$ = 66,$N_{GPT-4o}$ = 41) = 11.7, p-value = .205, i.e., there is no statistically significant difference between the severities of the issues identified by GPT-4o and experts.

\vspace{-0.3cm}
\subsection{Consolidated issues identified only by GPT-4o (RQ2)}

The results of the thematic analysis \cite{braun2024thematic} allowed us to classify the defects identified only by GPT-4o in five main categories: i) system visibility and feedback issues (D10, D23, D30, and D39); ii) navigation and learnability issues (D5, D18, D19, D24, and D28); iii) clarity in the interface and visual elements issues (D1, D11, D21, D31, and D35); iv) interactive elements and accessibility issues (D2, D16, D19, D32, D34, and D36); and v) context and irrelevant information issues (D22, D32, D33, D40, and D41).

Regarding \textit{system visibility and feedback issues}, GPT-4o identified flaws, such as the absence of a clear indicator for loading maps (D30), which can generate uncertainty for the user about whether the action was processed. A similar situation occurs in the lack of adequate visual feedback when users apply filters (D29), making it necessary for the user to investigate the interface to know whether a filter was activated. In addition, GPT-4o identified that the interface does not make it clear whether a search was successful or failed (D23), which can lead to unnecessary repeated attempts. GPT-4o also observed the absence of a visual indicator when drawing or measuring actions on the map (D39), challenging the user to perceive the progress of these operations. Furthermore, inadequate visibility of the system status at different times (D10) compromises the predictability of the interface, increasing the risk of confusion.

About \textit{navigation and learnability issues}, GPT-4o identified that the feature of changing the system's language (D5) does not clearly indicate possibility of interaction, introducing challenges to find and use it. GPT-4o found that the same occurs when navigating between filter categories (D18), where there is no clear visual distinction that certain elements are interactive, compromising the interface's intuitiveness. Another critical point highlighted by GPT-4o is the naming of buttons (D19), which are not self-explanatory. Some icons and labels do not clarify their purpose, requiring trial and error on the user's part. In addition, specific informational messages, such as "No active layer!" (D24), do not provide clear directions on how to proceed, which can generate doubts. Navigation between functionalities (D28) also lacks greater organization, as there is no clear way to switch between options without the need for repeated interactions.

Concerning \textit{clarity in the interface and visual elements issues}, GPT-4o identified an overload of information in the system's side panel (D1), which makes it challenging to see features and results in a visually dense interface. In addition, the combination of graphics on the map (D11) can introduce barriers for the understanding the visual hierarchy, compromising the understanding of which elements have priority in the interaction. GPT-4o also noted layout problems in the general structure of the interface (D21), which reduces intuitiveness and increases the cognitive effort required to find and interpret information. Furthermore, overlapping windows (D31) were observed, hindering the ability to view important content and complicating navigation. Lastly, GPT-4o suggested improvements to the visual index of the maps (D35), as the thumbnails do not clearly differentiate between the available options.

Regarding \textit{interactive elements and accessibility issues}, according to GPT-4o, the zoom buttons were considered too small for interaction (D16), which can compromise their usability, especially on touchscreen devices. In addition, their additional functionality is not indicated (D2), preventing the user from quickly understanding their options. For GPT-4o, other buttons in the interface also present problems, such as the "Clear all" button (D36), whose function is not well defined, which can lead to the accidental loss of information without prior confirmation. Likewise, the close pop-up button (D34) is small and difficult to view, which can hinder the fluidity of the interaction. The lack of clear indication of the coordinates displayed on the map (D32) was also identified by GPT-4o, making this information inaccessible to users who need it frequently.

About \textit{context and irrelevant information issues}, GPT-4o found that some content may be considered irrelevant to most users (D22), such as map credits, which take up space without any apparent need. Context issues were also identified, such as the "Road" label in English (D33), which can confuse users who expect a fully translated interface. In addition, the base maps do not clearly describe the type of information they represent (D40), making it difficult to understand their usefulness and application. GPT-4o also pointed out that the presentation of coordinates in the system presents two problems: the lack of an explicit indication of what they represent (D32) and the low visibility of this information (D41), which appears in a reduced size and may go unnoticed. These factors compromise the accessibility of the functionality for users who depend on this information for their activities.

\subsection{False positives identified by GPT-4o (RQ3)}

As previously indicated, 27 issues were classified as false positives and categorized as problem disagreement, problem assumptions, or generalized problems (see Section 4.2). Table \ref{tab:false} shows the false positives identified in each category.

\begin{table}[ht]
    \centering
    \caption{False positives (FP) identified by GPT-4o.}
    \scriptsize
    \begin{tabular}{ccl}\hline
     \textbf{Category} & \textbf{ID} & \textbf{Description} \\\hline
     Problem disagreements & FP1 & Inconsistency in the use of icons and text\\
     & FP2 & Contact links are not prominent\\
     & FP3 & Inconsistency in the design of the top and bottom menu\\
     & FP4 & Limited visibility of data in the graph\\
     & FP5 & Overlapping of buttons and information\\
     & FP6 & Small graph legend\\
     & FP7 & Map lacks interaction instructions\\
     & FP8 & Correspondence between the system and the real world\\
     & FP9 & Inconsistent terminology\\
     & FP10 & Menu options lack standardized iconography\\
     & FP11 & Icon consistency\\
     & FP12 & Recognition rather than recall\\
     & FP13 & Search option not very prominent\\
     & FP14 & Favorite icon is not self-explanatory\\
     & FP15 & Unclear mix of icons and text\\\hline
     Problem assumptions & FP16 & Lack of feedback on interactive elements\\
     & FP17 & Lack of clear error indication\\
     & FP18 & No feedback when clicking on sidebar items\\
     & FP19 & Lack of confirmation of actions\\
     & FP20 & Lack of feedback on actions\\
     & FP21 & Error prevention\\
     & FP22 & System status visibility\\
     & FP23 & Lack of user feedback\\
     & FP24 & Lack of direct feedback when hovering the mouse\\\hline
     Generalized problems & FP25 & Interface is overloaded with unnecessary information\\
     & FP26 & High cognitive load\\
     & FP27 & Flexibility and efficiency of use\\\hline
    \end{tabular}
    \label{tab:false}
\end{table}

The \textit{problem disagreements} category grouped issues identified by GPT-4o, but the researchers did not find them during the screenshot analysis. Many issues, such as inconsistency in the use of icons and texts (FP1), lack of standardization in the iconography (FP10), and lack of prominence of some elements (FP2, FP13), were refuted by the researchers after the screenshot inspections, as they simply did not exist. Other issues, such as unclear mix of icons and text (FP15) and a small graph legend (FP6), involve the design perceptions of GPT-4o and the researchers, who disagreed that these problems were real issues.

For \textit{problem assumptions} category, another recurring factor in GPT-4o was the difficulty in verifying specific issues in an analysis based only on static images. GPT-4o indicated the lack of feedback from the system (FP16) or the absence of action confirmations (FP19), but when testing the system, these elements were present. The lack of feedback in actions such as clicking on sidebar items (FP18) and hovering the mouse (FP24) were all identified as false positives because, in practice, the system offered adequate visual and interactive responses. In addition, some false positives arose due to an error in assessing confirmation messages before critical actions. Reports such as lack of clear error indication (FP17) were refuted when it was found that the system offered these warnings.

Taking into account the \textit{generalized problems} group, false positives were characterized by excessively generic evaluations that could not be verified. Issues such as interface overloaded with unnecessary information (FP25), high cognitive load (FP26), and lack of flexibility and efficiency of use (FP27) are examples of problems described broadly without sufficient detail to confirm their validity.

\vspace{-0.3cm}
\section{Discussion \& Takeaways}

By answering RQ1 \textbf{(What is the agreement between heuristic evaluation results by experts and the results obtained by GPT-4o, in terms of issue compliance, violated heuristics, and severity?)}, we can conclude that: i) few issues were identified by both groups (GPT-4o and human experts) and, those that were, they differed in terms of writing and detail; ii) both groups identified few issues related to heuristics of help and documentation, but GPT-4o did not identify any issues related to help users recognize, diagnose, and recover from errors; and iii) the severity tends to be the same for both groups, which are more likely to identify minor or major issues and few cosmetic or catastrophic.

We noticed that, despite pointing to the same problem, GPT-4o and experts differ in terms of violated heuristics and severity. Furthermore, the issues identified by both are closely linked to heuristics H2 (match between system and the real world) and H8 (aesthetics and minimalist design), the heuristics that allowed GPT-4o to identify issues the most and not the other way around. For example, H6 (recognition rather than recall), the heuristic that allowed experts to identify more issues, only allowed GPT-4o to identify one issue, which experts also identified. Furthermore, GPT-4o describes the issues more directly, while experts describe them in more detail, even using questions. We infer that experts had a greater argumentative, provocative, and explanatory capacity for issues. 

\begin{center}
\small
\fbox{\begin{minipage}{\dimexpr\textwidth-0.3cm}
\textbf{Takeaway 1.} GPT-4o and experts converge in finding some issues, but GPT-4o lacks depth, highlighting the gap between heuristic detection and human reasoning.
\end{minipage}}
\end{center}

Both groups found very few issues related to heuristics H9 (helping users recognize, diagnose, and recover from erros) and H10 (help and documentation), suggesting that the system does not really have many issues related to user help. In any case, our results were not promising for GPT-4o concerning these heuristics. This result could also be because there were no interactions during the evaluation performed by GPT-4o, since if there were, it could expand menus or search for documentation deeper within the system. We believe that more in-depth studies on these specific heuristics are needed to infer GPT-4o’s ability to identify issues related to them. Moreover, the low agreement between GPT-4o and the experts can generate some practical consequences, such as lack of confidence in the LLM, misalignment in usability priorities (GPT-4o identifying issues that are not so critical) and more workload for experts/evaluators since they will always need to review and evaluate the results generated by the LLM.

About severity, our results also corroborate those of Meinecke et al. \cite{meinecke2024comparative}, suggesting that there is no difference between the severity ratings of issues identified by GPT-4o and by experts. Most issues classified by GPT-4o and experts were severity 2 (minor issue) or 3 (major issue). This result may be linked to the fact that the system does not have catastrophic or cosmetic issues. However, this may also be connected to the neutral response bias, suggesting that groups tend to choose an intermediate option on a scale instead of the most extreme options \cite{edwards2014effects,nowlis2002coping}. Some studies \cite{nowlis2002coping,lewis2023} suggest that the absence of an intermediate option on scales tends to increase the choice of other options by a small amount. In our case, we do not suggest removing the severity scales widely validated in the literature to identify issues in general in systems, but rather to conduct further studies to identify how GPT-4o behaves to identify only cosmetic or catastrophic issues in a selected sample. 

\begin{center}
\small
\fbox{\begin{minipage}{\dimexpr\textwidth-0.3cm}
\textbf{Takeaway 2.} GPT-4o assigns severity ratings similar to experts, identifying mainly minor and major issues according to the levels proposed by heuristic evaluation.
\end{minipage}}
\end{center}

Considering the RQ2 \textbf{(What are the characteristics of the issues found exclusively by GPT-4o?)}, we highlight the following conclusions: i) GPT-4o tends to identify issues based on design patterns; ii) GPT-4o identifies more issues related to match between the system and the real world (H2) and aesthetics and minimalist design (H8), but loses in issues related to control, efficiency, and flexibility (H3 and H7).

We noticed that the experts tended to focus on the real context of use and user interaction. In contrast, GPT-4o focused on general heuristics and patterns followed. Our results showed, for instance, that GPT-4o identified information overload in the side panel (see D1 in Table \ref{tab:issues}), indicating that excess options can saturate the user and make decision-making difficult. Experts may have considered this information acceptable or necessary because they were already used to navigated or inspect complex systems. Another example of this difference in issues between the groups is the arrangement of elements on the screen, such as the overlapping of windows (see D31 in Table \ref{tab:issues}) and the confusion generated by the combination of graphs and maps (see D11 in Table \ref{tab:issues}). GPT-4o may have perceived these issues more clearly, basing its analyses on general visual design principles, while experts may have automatically adapted to the interface and its arrangement. These results align with those of Meinecke et al. \cite{meinecke2024comparative}, who mention that GPT-4o does not necessarily identify issues with more validity than human evaluators. In fact, our study suggests that GPT-4o can complement the experts' evaluation by identifying potentially different issues.

\begin{center}
\small
\fbox{\begin{minipage}{\dimexpr\textwidth-0.3cm}
\textbf{Takeaway 3.} GPT-4o complements expert evaluations by identifying heuristic-based issues that experts might overlook due to familiarity with complex systems.
\end{minipage}}
\end{center}

A possible hybrid approach to heuristic evaluation combining GPT-4o automation with human validation could involve using an LLM in a pre-analysis to identify potential problems, followed by human analysis of the findings to filter out inconsistent detections. Then, experts can review the problems detected by LLM and perform the evaluation focusing on other heuristics that the LLM has not explored properly. Nevertheless, this process needs to be tested and validated to confirm its results.

Digging deeper into the issues and violated heuristics, GPT-4o identified more issues for heuristics H2 (match between system and the real world) and H8 (aesthetic and minimalist design) than experts. This result corroborates the difference in issues shown in the discussion of the previous result since experts identified more issues related to H3 (user control and freedom) and H7 (flexibility and efficiency of use). Considering that experts are usually humans with extensive experience in HCI, we understand that
they are more apt to identify issues involving user interaction, flexibility, control, and freedom. However, this result may be linked to the fact that GPT-4o analyzed screenshots of the system; that is, it did not interact with elements on the screen. 

\begin{center}
\small
\fbox{\begin{minipage}{\dimexpr\textwidth-0.3cm}
\textbf{Takeaway 4.} GPT-4o can stand out among experts in identifying issues related to the match between the system and the real world and to aesthetic and minimalist design when it analyzes screenshots. However, it is outperformed on flexibility and user control.
\end{minipage}}
\end{center}

Concerning our RQ3 \textbf{(What are the characteristics of issues considered false positives identified by GPT-4o?)}, we initially highlight that 24.3\% of issues identified were false positives. In this sense, we can conclude the following: i) GPT-4o can also hallucinate to the identification of issues, finding issues where there are none; ii) GPT-4o tried to predict issues, simulating an interaction context even when analyzing screenshots; iii) GPT-4o may be too generic in the description of the issue.

Hallucination is the term given to a phenomenon that occurs in systems that use natural language generation, such as GPT-4o, when they provide outputs that seem faithful but, upon closer analysis, prove to be meaningless \cite{alkaissi2023artificial,ji2023survey}. In this context, we infer that in our study, GPT-4o may have hallucinated since it pointed out several issues that, when we checked, did not exist in the analyzed screenshot. Another hypothesis for identifying these false positives is GPT-4o's ability to assume issues in the system. We noticed that, even without interacting with the system, GPT-4o pointed out issues that could only be identified using the system, such as lack of feedback when hovering the mouse (FP24) and lack of feedback when clicking on a sidebar item (FP18). 

\begin{center}
\small

\fbox{\begin{minipage}{\dimexpr\textwidth-0.3cm}
\textbf{Takeaway 5.} GPT-4o can hallucinate or try to predict issues that do not exist, generating false positives. Human-in-the-loop is needed to validate the issues.
\end{minipage}}
\end{center}

Furthermore, there are also issues that are too general and do not provide enough information to understand the issue. Although there are few false positives in this category, we believe that modifying the prompt to request concrete and specific examples of the issues is a solution to avoid this result. 

Relying on LLMs for heuristic evaluation can result in false positives, the need for human validation, overgeneralization, and difficulties in identifying problems of interaction with the interface. Although these models can help detect many issues, they may point out non-existent issues or generically describe problems, requiring careful manual review. Since they do not interact directly with the system, they may erroneously infer usability issues as we showed. Therefore, LLMs should be used as a support, and not as the sole tool, in heuristic evaluation.

\vspace{-0.3cm}
\subsection{Limitations}

While the model generally behaved reliably and produced consistent responses, some hallucinations were observed, making occasional inaccuracies an inherent limitation of using an LLM for defect identification. We mitigated this by analyzing and filtering outputs, but some inconsistencies remain. Additionally, we only used OpenAI’s gpt-4o-2024-08-06 as it was one of the leading LLMs at the time of data collection (November 2024). However, future versions may yield different results, even when applying the same prompt, since the logic of these models differs from each other and evolves. While this limits the generalizability of our findings, we understand that experimenting with a leading model presents valuable insights concerning LLMs’ potential in heuristic evaluation. While multiple researchers validated the defect identification process, disagreements were resolved through discussion, introducing an element of subjectivity, particularly in assessing defect severity and relevance. Moreover, despite detailing our methodology and making study artifacts available, the evolving nature of LLMs may impact the replicability and consistency of results over time. Finally, another limitation of our study is the generalization of the results to other types of software. In our study, we evaluated only one web system, and our results can be generalized to this type of system. More studies are required to generalize and replicate the results to other systems and contexts, such as mobile applications.

\vspace{-0.1cm}
\section{Conclusion}

We highlight three main contributions based on the results and discussions presented in the study: i) we developed, tested, and made available a prompt to be used by HCI researchers who want to perform heuristic evaluations using GPT-4o and screenshots. To the best of our knowledge, our paper is the first to perform heuristic evaluation on GPT-4o using screenshots of a web-based system and provide a literature-grounded prompt with prompt engineering elements; ii) we described a complete process for classifying issues identified by GPT-4o, including duplicate removal and false positive classification; iii) we provide five takeaways that can foster the conscious use of GPT-4o in heuristic evaluations. Our results indicate that GPT-4o should not be treated as a replacement for human experts but rather as a complement. 

As future work, researchers can explore the ability of GPT-4o and prompt engineering to generate issues for specific heuristics, for example, those related to user flexibility and efficiency, and help heuristics. It is also interesting to verify the behavior of LLMs in identifying issues of more extreme severity, such as cosmetic and catastrophic issues. Furthermore, researchers can also conduct more in-depth investigations on prompt engineering and heuristic evaluation, testing variations and prompt styles in the same interfaces to compare the findings. Lastly, despite GPT-4o's relevance, future research could explore other LLMs, which might differ on the generated output and the resources required to use. 

\section*{Acknowledgments}

We thank the partial financial support of Conselho Nacional de Desenvolvimento Científico e Tecnológico (CNPq - Brazil) - grant 309497/2022-1, Coordenação de Aperfeiçoamento de Pessoal de Nível Superior (CAPES - Brazil) - Finance Code 001, and Fundação Araucária de Apoio ao Desenvolvimento Científico e Tecnológico do Estado do Paraná (FA - Brazil).

 \bibliographystyle{splncs04}
 \bibliography{samplepaper}

\end{document}